\documentstyle[aps,prl,twocolumn]{revtex}
\begin{document}
\draft
\title{Dephasing of Electrons on Helium by Collisions with Gas Atoms}
\author{D. Herman, H. Mathur and A.J. Dahm}
\address{Department of Physics, Case Western Reserve University,
Cleveland, OH 44106-7079} 


\date{today}
\maketitle
\begin{abstract}
The damping of quantum effects in the transport properties
of electrons deposited on a surface of liquid helium is
studied. It is found that due to vertical motion of the
helium vapour atoms the interference of paths of duration
$t$ is damped by a factor $\exp - (t/\tau_v)^3$. An expression
is derived for the weak-localization lineshape in
the case that damping occurs by a combination of processes 
with this type of cubic exponential damping and processes with
a simple exponential damping factor. 
\end{abstract}
\pacs{PACS: }
\input epsf

\section{Introduction}

The damping of quantum effects in a system coupled to external
degrees of freedom is a fundamental problem of atomic physics,
condensed matter physics and quantum optics. There is great 
interest in understanding and controlling such damping in well
characterised systems. Here we study the damping of quantum
effects in the transport properties of a two-dimensional electron
gas deposited on the surface of a pool of liquid helium \cite{review}. 

Electrons on the surface of helium are vertically confined by 
their image charges and an (optional) applied holding field. 
They constitute a two-dimensional electron gas 
similar to those in semiconductor devices \cite{ando} but with 
different scattering and damping mechanisms. Electrons may 
scatter off ripples on the surface of the helium pool (``ripplons'')
or off helium vapour atoms above the liquid surface. Above 1 K,
gas atom scattering dominates \cite{review}, and we concentrate 
on this regime. 
On the electronic time-scale, the helium vapour atoms are almost
stationary and hence similar to impurities in a metal film
or a semiconductor device. Thus there are quantum interference
corrections to the resistivity at low temperature familiar from
studies of transport in metals and semiconductors \cite{lee}. 
These corrections
result from constructive interference between closed electron
paths and their time-reversed counterparts, leading to a small
enhancement of the resistance (``weak-localization correction'').

The slow movement of helium atoms leads to damping of weak-localization.
There is an important distinction between the effect of vertical 
and horizontal motion of helium atoms. Roughly, horizontal 
movement produces damping by scrambling the phase of the interfering
paths; vertical movement, by reducing the weight of contributing
paths of long duration. The effect of horizontal movement has
been analysed previously \cite{afonin}; it is the purpose of this paper to 
study the effect of vertical motion.

The central result is that due to vertical motion of the
helium atoms the interference contribution of paths of
duration $t$ is reduced by a factor $\exp - (t/\tau_{v})^3 $.
Thus paths of duration greater than the damping time, $\tau_v$,
are effectively cutoff. An interesting feature is that damping
due to both vertical and horizontal movement of helium atoms
is not a simple exponential; it cuts off more sharply as the
exponential of $t^3$. In contrast, electron-electron and
electron-phonon interactions in metals and semiconductors
are supposed to produce simple exponential damping. Damping
in atomic physics and nuclear magnetic resonance is also
commonly a simple exponential; this is indicated by the
Lorentzian shape of spectral and magnetic resonance 
lines \cite{atom,nmr}\footnote{Recall that the Fourier transform
of $e^{-|t|}$ is a Lorentzian.}. As emphasized by Afonin
{\em et al.} \cite{afonin} in context of quantum transport, 
the form of damping can be probed by measuring the
magnetic field dependence of the weak-localization correction
(``weak-localization lineshape''). In section III we exhibit
some lineshapes corresponding to different forms of damping.

Weak-localization has been observed in a related system, electrons
on a surface of solid hydrogen \cite{adams}. In this system helium vapour
was deliberately introduced above the solid hydrogen to scatter
electrons; thus gas atom damping is relevant to this type of
experiment. More recently, Karakurt {\em et al.} have systematically
studied the dependence of the damping rate on various experimental
parameters (electron density; gas vapour pressure, controlled via
temperature; and holding field) for electrons on helium \cite{karakurt}. 
In this way they have obtained
quantitative information on the contributions of different 
mechanisms to the damping rate. It is the experiment of 
Karakurt {\em et al.} that prompted us to carry out the present
investigation.

For orientation it is useful to recall some typical parameters
for the experiment of Karakurt {\em et al.} In the absence of
a holding field, the electron is bound to the surface by its
image. The charge of the image is reduced from the bare charge
of the electron by a factor $ (\epsilon - 1)/(\epsilon + 1)
= 7 \times 10^{-3} $ \cite{jackson}; 
thus the vertical scale of the electronic wavefunction is 
76 ${\rm \AA}$. The lowest vertical subband wavefunction
is of the Fang-Howard form, $ \phi(z) \propto z 
\exp( - z/b )$, at zero holding field; this form
remains an excellent variational ansatz with $b$ an adjustable 
parameter when a holding field is applied. Here $z$ denotes
the distance of the electron above the helium pool. The subband spacing
is 6 K; hence for sufficiently low temperatures\footnote{At zero
field for an ideal interface the subband spectrum is like that of
atomic hydrogen. To calculate the ground state occupancy for such a spectrum 
it is neccessary to regulate the partition function
as discussed by Fermi\cite{fermi}. Hence estimation of the temperature
threshold below which the surface electrons are effectively two-dimensional
involves some subtlety \cite{review}.} and electron densities
below $2 \times 10^{15}$ 
/m$^2$ the surface electrons behave like a two-dimensional
electron gas. Much of the data of Karakurt {\em et al.} is 
at temperatures around 2 K and at a typical density of 
$2 \times 10^{11}$ /m$^2$ corresponding to a Fermi temperature
of 0.6 mK. Note
that their two-dimensional electron gas is therefore non-degenerate
in contrast to the situation in metal films and typical 
semiconductor devices. Thus transport properties are not determined
entirely by mono-energetic electrons on the Fermi surface; instead
we must sum the Boltzman-weighted contribution of electrons of all
energies. The electron-atom collision time inferred from mobility
measurements was typically a few ps. The longest relevant electronic
time scale is $\tau_z$, the time taken by a thermal electron to
move a distance $b$ (see eq 5 below). At 2 K and zero holding
field $\tau_z = 80$ ps.
In comparison, the atom-atom collision time is enormous,
of the order of 10 ns.

\section{Analysis of Damping}

In this section we analyse the damping produced by the vertical
motion of helium vapour atoms. First we analyse a simple model
(model I) that captures some of the essential physics, but leads to the
incorrect conclusion that the damping factor goes as the
exponential of $t^2$ rather than $t^3$, a result obtained earlier
by Stephen \cite{stephen}. We then identify a shortcoming of 
model I and in the next subsection introduce and analyse an
improved version (model II) that leads to the correct answer.

\subsection{Model I}

In this model we assume that the Helium atoms are able to scatter
electrons only if they are within a certain distance (denoted $b$)
from the
liquid helium surface. It is also assumed that the scattering is
independent of the precise height of the atom so long as it lies
within the prescribed distance.

Consider $p(t) = $ probability that an atom will remain within the
scattering distance for a time $t$. At first let us assume
the probability decays exponentially,
\begin{equation}
p(t) = \exp \left( - \frac{ t }{ \sqrt{\pi} \tau_z } \right);
\end{equation}
the numerical coefficient in the exponential is for later convenience.
Since the motion of vapour atoms is essentially independent
the probability that $n$ atoms will remain within the scattering
distance for time $t$ is 
\begin{equation}
[p(t)]^n = \exp \left( - \frac{ n t }{ \sqrt{\pi} \tau_z } \right).
\end{equation}

Weak-localization results from constructive interference between
the history in which an electron traverses a particular closed
path and the history in which it traverses the same path backwards.
A path of duration $t$ involves $t/\tau_e$ collisions. For this
path to contribute to weak-localization it is neccessary for all atoms
to remain within the electron-scattering region for a duration of order
$t$. Hence the fraction of paths of duration $t$ that contribute to 
weak-localization is 
\begin{equation}
\gamma (t) = \exp \left(  - \frac{t^2}{ \sqrt{\pi} \tau_e \tau_z} \right).
\end{equation}
Here $\gamma$ is the damping factor; essentially this result 
is given in ref \cite{stephen}.

This argument must be improved in two ways. 
First, it is not neccessary for all the atoms to remain in
place for the entire duration of a closed path. In particular,
atoms encountered by the electron in the middle of a closed
path are encountered by both the forward and backward path at
essentially the same time. At the other extreme, atoms encountered
early on the forward path are encountered towards the end on the backward
path, a time $t$ later. Thus on average atoms need to remain in
place for a time $t/2$. Hence the damping factor is really
\begin{eqnarray}
\gamma (t) & = & \exp \left( - \frac{t^2}{2 \sqrt{\pi} \tau_e \tau_z} \right) 
\nonumber \\
& \equiv & \exp \left( - \frac{ t^2 }{ \tau_v^2 } \right)
\end{eqnarray}
with $\tau_v = \pi^{1/4} \sqrt{ 2 \tau_e \tau_z /c }$.

A second improvement is needed because eq (1) is incorrect. 
$p(t)$ is easily calculated and seen to not be
exponential. Here we mention only the relevant features of $p(t)$;
the details are relegated to appendix A.
(i) As expected on dimensional grounds, $p(t)$ is a function of 
$t/\tau_z$ alone, where 
\begin{equation}
\tau_z  = \sqrt{ \frac{M b^2}{2 k T} }.
\end{equation}
Here $M =$ mass of a helium atom. Physically, $\tau_z$ is
the time taken by a thermal atom to move a distance $b$.
(ii) For short times, $ t \ll \tau_z$, we find
\begin{equation}
p(t) \approx 1 - \frac{t}{\sqrt{\pi} \tau_z}
\end{equation}
(iii) For long times, $t \gg \tau_z$, $p(t)$ vanishes in
a manner not relevant to our purpose.

Now the probability that $n$ atoms remain near the surface
is 
\begin{eqnarray}
[ p(t) ]^n & = & \exp [ n \ln p(t) ] \nonumber \\
 & = & \exp \left[ n \ln \left( 1 - \frac{t}{\sqrt{\pi}\tau_z} + \ldots \right) 
\right] \nonumber \\
& \approx & \exp \left( - \frac{ n t }{\sqrt{\pi} \tau_z} \right).
\end{eqnarray}
This shows that for large $n$, $[ p(t) ]^n $ can be approximated
as an exponential only for $ t \ll \tau_z/\sqrt{n} $; but since
it becomes negligible in any case once $ t \gg \tau_z/ n$, there
is no significant error in taking $ [p(t)]^n$ to be an exponential.

The upshot of this discussion is that although $p(t)$ is far from
exponential, $[p(t)]^n$ is a simple exponential under appropriate
circumstances; eq (2) is valid, although eq (1) is not. Similarly
we see that eq (4) is also valid provided $ \tau_z \gg \tau_e $, a 
condition needed for weak-localization. 

In summary, for model I the damping decays as the exponential of
$t^2$. Provided $\tau_z \gg \tau_e$, it is given by eq (4). The atomic
time constant $\tau_z$ is given by eq (5). Evidently, the three time
scales are arranged in the hierarchy $\tau_z > \tau_v > \tau_e$.

\subsection{Model II}

The shortcoming of model I is the assumption stated in the first
paragraph of the previous subsection. It is more realistic to
assume that the ability of an atom to scatter electrons turns off
smoothly as it moves away from the liquid helium surface. 

If we treat the atoms as hard core potentials, the contribution of
a closed path to the return amplitude is a product of the amplitude for
the electron to go to atom 1, multiplied by the amplitude to scatter
off atom 1, multiplied by the amplitude to go to atom 2, multiplied
by the amplitude to scatter off atom 2, and so on around the loop.

Let $A(z)$ be the amplitude to scatter from an atom at height $z$
above the helium surface. Model I can be described as the case
in which $A(z)$ is a step function. Here we choose
\begin{eqnarray}
A(z) & = & \frac{4 \lambda z^2}{b^3} \exp \left( - \frac{ 2 z }{b} \right) 
\hspace{3mm} {\rm for} \hspace{2mm} z > 0; \nonumber \\
& = & 0 \hspace{5mm} {\rm for} \hspace{2mm} z < 0.
\end{eqnarray}
This is derived by taking the vertical subband wavefunction of
the electrons to be of the Fang-Howard form and treating the
helium atom as a short-ranged hard-core potential.

If the helium atoms are only allowed to move vertically
the forward and backward paths remain in phase; however
the interference contribution to the return probability
is still modified because the forward and backward paths
have different amplitudes to scatter from each atom.
We must consider
\begin{equation}
Q(t) = \langle A(z) A(z+vt) \rangle.
\end{equation}
Here $t$ is the difference in the times at which the atom
is encountered on the forward and return path. The atom is
assumed to move ballistically at vertical speed $v$ for this time.
$\langle \ldots \rangle$ denotes an average over all possible
configurations of the Helium atom (vertical position is assumed
to be uniformly distributed and vertical speed is given by the
Maxwell-Boltzmann formula). 

Introduce the normalization factor $R(t)$ defined by
\begin{equation}
R(t) = \langle A(z) A(z+vt) \rangle.
\end{equation}
Here the average over vertical position is performed as in eq (9)
but the velocity distribution is assumed to be a delta function
peaked about zero. $R(t)$ is the value of $Q(t)$ when the atoms
don't move. Let 
\begin{equation}
q(t) = Q(t)/R(t).
\end{equation}
The contribution of paths of duration $t$ is then reduced roughly
by the factor $q(t)$ raised to the power $t/\tau_e$, the number of
atoms encountered. 

$q(t)$ is analogous to $p(t)$ for model I. Again on dimensional
grounds, $q(t)$ depends only on the ratio $t/\tau_z$ and
again we are interested only in the short time behaviour.
This is evaluated in Appendix B. The difference from the
previous case is that
\begin{equation}
q(t) = 1 - \frac{ t^2 }{3 \tau_z^2 } + \ldots
\end{equation}
for short times, $t \ll \tau_z$. The behaviour is quadratic
rather than linear (compare eq 6). 
Quadratic behaviour is generic; the linear behaviour for model I
is an artifact of the discontinuous step in $A(z)$. 
Hence $q(t)$ raised to the power
of $n$ is approximately Gaussian rather than exponential
\begin{equation}
[ q(t) ]^n \approx \exp \left( - \frac{n t^2 }{3 \tau_z^2 }
\right).
\end{equation}
Eq (13) should be contrasted with eq (7) above for model I.

To obtain the damping factor, roughly we must replace 
$n$ in eq (13) by $t/\tau_e$, the number of atoms encountered
in a path of duration $t$. Before that we must replace $t^2$
in eq (13) by $t^2/3$, its value averaged over the interval
from 0 to $t$ with uniform weight. This is to take into account
the range in the difference of times at which an atom is
encountered along the forward and reversed histories.

The result for the damping factor is
\begin{equation}
\gamma (t) = \exp \left( - \frac{ t^3 }{\tau_v^3} \right)
\end{equation}
where $ \tau_v = (9 \tau_e \tau_z^2)^{1/3}$. 
Eq (14) is the central result of this
paper. It is valid provided $ \tau_z \gg \tau_e$. 

\section{Lineshape}

Karakurt {\em et al.} observed damping by vapour atom motion
at low electron density and by electron-electron interaction 
at high density \cite{karakurt}. 
At intermediate densities, damping by both
mechanisms was substantial. Vapour atom scattering produces
cubic exponential damping; electron-electron interaction is
presumably simple exponential. Afonin {\em et al.} have pointed
out that the weak-localization lineshape depends on the form
of damping and they have given an expression for the lineshape
in the extreme cases that the damping is entirely simple
exponential or entirely cubic exponential. The purpose of this
section is to study the lineshape in the intermediate regime
and examine how it crosses over from one extreme form to the 
other.

For simplicity, first let us consider a degenerate electronic
system. Assuming that the different damping mechanisms are
independent the lineshape is given by
\begin{eqnarray}
\delta g (E, B) & = & - \frac{1}{\pi} \frac{e^2}{h} 
\left( \frac{W}{L} \right) \phi (E, B);
\nonumber \\
\phi(E,B) & = & \int_{\tau_e}^{\infty} d t \frac{ 4 \pi e D B}{h} 
\frac{ e^{ - t/\tau_1 } e^{ - t^3/\tau_3^3 } }{\sinh ( 4 \pi e D t B/h )}.
\end{eqnarray}
Here $E$ is the Fermi energy; $W$, the sample width;
$L$, the sample length; $D$ the electron diffusion 
constant; $1/\tau_1$, the simple exponential damping
rate; and $1/\tau_3$, the cubic exponential damping
rate. Energy dependence enters the integrand in eq (15)
through the diffusion constant $D = E \tau_e/m$ and through
the energy dependence (if any) of the time constants $\tau_1$
and $\tau_3$. The $\sinh$ factor in eq (15) may be recognised
as the Fourier transform of the directed area distribution
for closed random walks on a plane \cite{harold}. 

It is useful to manipulate eq (15) into a more revealing form.
To this end introduce the dimensionless variable $ u = 8 \pi e D t B/h$
to obtain
\begin{equation}
\delta g = - \frac{1}{\pi} \frac{e^2}{h} 
\left( \frac{W}{L} \right)
\int_{B/B_e}^{\infty} d u \frac{ \exp - u \left( \frac{1}{2} 
+ \frac{B_1}{B} \right) \exp \left( - u^3 \frac{B_3^3}{B^3} 
\right)}{1 - e^{-u}};
\end{equation}
here $B_e = h/(8 \pi e D \tau_e)$. Making use of the asymptotic
formula
\begin{equation}
\int_{\epsilon}^{\infty} d u \frac{ e^{-u} }{u} \approx \ln \frac{1}{\epsilon}
+ \gamma
\end{equation}
we obtain
\begin{eqnarray}
\delta g / \left( \frac{e^2}{h} \frac{W}{L} \right) 
& = & - \frac{1}{2 \pi}
\left[ \ln \frac{B_e}{B_1} + \ln \frac{B_e}{B_3} \right] \nonumber \\
 & & + \frac{1}{\pi} {\cal F} \left( \frac{B_1}{B}, \frac{B_3}{B} 
\right)
\end{eqnarray}
where $B_1 = h/(8 \pi e D \tau_1)$, $B_2 = h/(8 \pi e D \tau_3)$,
$\gamma = 0.577216 \ldots$ is Euler's constant and the function
\begin{eqnarray}
{\cal F}(x,y) & = & \frac{1}{2} \ln x + \frac{1}{2} \ln y + \gamma
\nonumber \\
& & 
- \int_{0}^{\infty} d u \left[ 
\frac{e^{-u}}{u} - \frac{ \exp - u \left( \frac{1}{2} + x \right)
\exp \left( - u^3 y^3 \right) }{ 1 - e^{-u} } \right].
\nonumber \\
& &
\end{eqnarray}
Eqs (18) and (19) constitute the generalisation of the standard
weak-localisation lineshape to the case that both $\tau_1$ and
$\tau_3$ damping are present. For the special case that there 
is no $\tau_3$ damping (hence $y \rightarrow 0$) eqs (18,19) 
reduce to the familiar expression involving digamma functions
by use of the integral representation \cite{morse}
\begin{equation}
\int_{0}^{\infty} d u \left( \frac{ e^{-u} }{u} -
\frac{ e^{- \left( \frac{1}{2} + x \right) u }}{ 1 - e^{-u} }
\right) = \psi ( \frac{1}{2} + x ).
\end{equation}
A significant feature revealed by eqs (18,19) is that the
lineshape is universal: ${\cal F}$ does not depend on microscopic
length scales. Note that the magnetic field dependence is entirely
in the second term of eq (18); the first term is an
additive constant. A practical advantage of eq (19) over eq (15)
is that the integrand is well behaved for both large and small
$u$. In contrast, the integrand in eq (15) diverges at the
lower end.

To study the crossover in lineshape we fix the damping rate
$ 1/\tau_1 + 1/\tau_3 =
1/\tau_{\phi} $. Equivalently, we fix $B_1 + B_3 = B_{\phi}$. $\delta g$ is
plotted as a function of $B$ for several values of the ratio $B_1/B_{\phi}$.
Fig 1 shows that for the same damping rate the lineshape changes noticeably
as damping shifts from simple exponential to cubic exponential.

Fig 2 shows the behaviour of the conductance minimum at $B=0$ for
a fixed damping rate. It is given by
\begin{equation}
\delta g (B=0) = - \frac{1}{\pi} \frac{e^2}{h} 
\left[ \ln \left( \frac{B_e}{B_{\phi}} \right) +
u \left( \frac{B_3}{B_1} \right) \right]
\end{equation}
with the crossover function
\begin{equation}
u(x) = \ln (1 + x) + \gamma + \int_{0}^{\infty} d s
(1 + 3 s^2 x^3) \ln s e^{-s} e^{-s^3 x^3}.
\end{equation}
As implied by eqs (18) and (19) the crossover depends only
on the ratio $B_3/B_1$. $u$ has the limiting values
$u(0) = 0$ and $u(\infty) = 2 \gamma/3$.

Under experimental conditions \cite{adams,karakurt} the electron
gas is non-degenerate. At finite temperature
\begin{eqnarray}
\delta g (T,B) & = &
- \int_{0}^{\infty} d E \frac{\partial f}{\partial E} \delta g (E,B)
\nonumber \\
& \approx & \frac{n \pi \hbar^2}{m (kT)^2} \int_{E_c}^{\infty}
d E \delta g(E,B) e^{- E/kT}.
\end{eqnarray}
$n$ is the area density of electrons.
In the second line of eq (23) we have approximated the Fermi
function by a Boltzman factor and imposed a lower cutoff $E_c$. Below
the cutoff energy the electrons are presumed to be strongly 
localized and to make an insignificant contribution to the 
conductance. These finite temperature considerations make
it more difficult to extract the form of damping from the
lineshape. 

\section{Conclusion}

In summary, we have given a physical argument that due to 
vertical motion of helium atoms the interference of electron
paths of duration $t$ is damped by a factor $\exp - (t/\tau_v)^3$.
We have derived a formula for the universal magnetoconductance 
lineshape for the case that both $\tau_1$ and $\tau_3$ damping
are present. It should be possible to rederive these results
via impurity averaged diagrams; this is left open for future
work.

It is a pleasure to acknowledge helpful correspondence
with M. Stephen. 
This work was supported in part by NSF Grants
DMR 98-04983 (DH and HM) and DMR 97-01428 (AJD) 
and by the Alfred P. Sloan Foundation (HM).
HM acknowledges the hospitality of the Aspen Center for
Physics where this work was completed.

\appendix

\section{Asymptotics of $p(t)$}

We wish to calculate $p(t)$, the probability that a
vapour atom will remain within a vertical elevation $b$
of the liquid surface for a time $t$. We assume (i) the initial
elevation of the atom is uniformly distributed between zero
and $b$; (ii) the vertical velocity is Maxwell-Boltzman distributed;
(iii) the atom moves ballistically; and (iv) if the atom strikes
the liquid surface it sticks and does not reflect \cite{stick}.
Due to assumption (iii) the expression for $p(t)$ that we derive 
is valid only for times short compared to the atom-atom collision
time; however this is not a serious restriction since we are 
interested only in the short time behaviour of $p(t)$. 

Based on these assumptions we may write
\begin{eqnarray}
p(t) & = & \int_{0}^{b} d z \frac{1}{b} \int_{0}^{(b-z)/t} d v
\sqrt{ \frac{M}{2 \pi k T} } \exp \left( - \frac{M v^2}{2 k T} 
\right) \nonumber \\
 &  & + \int_{0}^{b} d z \frac{1}{b} \int_{-z/t}^{0} d v
\sqrt{ \frac{M}{2 \pi k T} } \exp \left( - \frac{M v^2}{2 k T} 
\right).
\end{eqnarray}
The two contributions correspond to the atom moving up and
down respectively.

By exchanging the order of integration we can perform the
$z$ integral first to obtain
\begin{equation}
p(t) = \frac{2}{\sqrt{\pi}} \int_{0}^{1/\overline{t}} d u
(1 - u \overline{t}) \exp( - u^2 ).
\end{equation}
We have rescaled variables so that $ u = v/\sqrt{2 k T/M }$
and $ \overline{t} = t/\tau_z $. Note that $p(t=0) = 1$
and as $ t \rightarrow \infty$, $p(t) \rightarrow 0$. 
Eq (A2) is an exact expression for $p(t)$. The small time,
$t \ll \tau_z$, asymptotic behaviour is
\begin{equation}
p(t) \approx \left( 1 - \frac{\overline{t}}{\sqrt{\pi}} + \ldots \right).
\end{equation}

\section{Asymptotics of $q(t)$}

To calculate $q(t)$ we assume that the initial elevation of the
vapour atom is uniformly distributed between the liquid surface
and an upper cutoff $L$. Ultimately we shall take $ L \rightarrow
\infty$. Aside from this we share the assumptions (ii), (iii) and
(iv) of Appendix A. 

Hence we obtain
\begin{eqnarray}
Q(t) & = & \langle A(z) A(z + v t) \rangle
\nonumber \\
& = & \frac{1}{L} \sqrt{ \frac{M}{2 \pi k T} }
\int_{0}^{L} d z \int_{-\infty}^{\infty} d v 
\exp \left( - \frac{M v^2}{2 k T} \right)
\langle A(z) A(z + v t) \rangle. \nonumber \\
 & &
\end{eqnarray}
Using eq (8) for $A(z)$ and rescaling we obtain
\begin{eqnarray}
Q(t) & = & \frac{16}{\sqrt{\pi}} \frac{ \lambda^2 }{ b L }
\int_{0}^{\infty} d u e^{-u^2} e^{- 2 u \overline{t} }
\int_{0}^{L/b} d \zeta e^{-4 \zeta} \zeta^2 (\zeta + u \overline{t})^2
\nonumber \\
 & & + \frac{16}{\sqrt{\pi}} \frac{ \lambda^2 }{ b L }
\int_{-\infty}^{0} d u e^{-u^2} e^{- 2 u \overline{t} }
\int_{- u \overline{t}}^{L/b} 
d \zeta e^{-4 \zeta} \zeta^2 (\zeta + u \overline{t})^2
\nonumber \\
& &
\end{eqnarray}
Here $ \overline{t} = t/\tau_z$, $u = v/\sqrt{ (2 k T)/M }$
and $\zeta = z/b$.

Performing the $\zeta$ integral yields
\begin{equation}
Q(t) = \frac{ \lambda^2 }{ 4 \sqrt{\pi} b L }
\int_{0}^{\infty} d u
e^{-u^2} e^{- 2 u \overline{t} }
(3 + 6 u \overline{t} + 4 u^2 \overline{t}^2)
\end{equation}
and hence the normalization
\begin{equation}
R(t) = Q(0) = \frac{3 \lambda^2 }{ 8 b L }.
\end{equation}

The exact reduction factor is then
\begin{eqnarray}
q(t) & = & \frac{Q(t)}{R(t)} \nonumber \\
& = & \frac{2}{3 \sqrt{\pi}} \int_{0}^{\infty}
d u e^{-u^2} e^{-2 u \overline{t}}
(3 + 6 u \overline{t} + 4 u^2 \overline{t}^2 )
\end{eqnarray}
with the small time, $ t \ll \tau_z $, asymptotic
behaviour
\begin{equation}
q(t) \approx 1 - \frac{ \overline{t}^2 }{3}
+ \frac{ \overline{t}^4 }{2} 
+ \ldots
\end{equation}


\newpage

\epsfxsize=3.0in \epsfbox{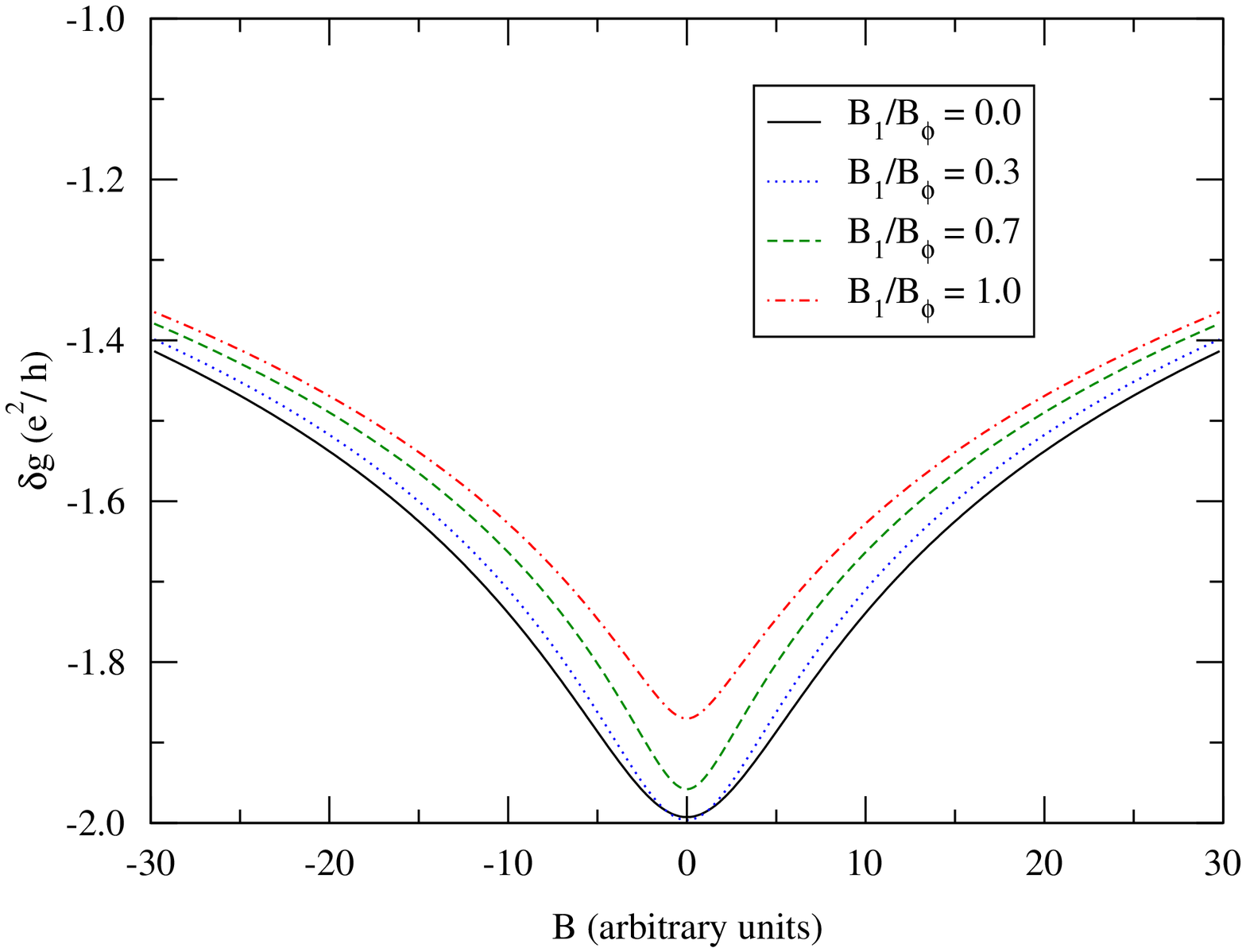}
\figure{ Figure 1. {\em Lineshape crossover:} The conductance
of degenerate electrons is plotted in units of $e^2/h$ as a
function of magnetic field.
The damping rate is held fixed ($B_1 + B_3 = B_{\phi}$).
Different curves correspond to different proportions of $\tau_1$
and $\tau_3$ damping, measured by the ratio $B_1/B_{\phi}$.
$B_1/B_{\phi} = 1$ corresponds to pure $\tau_1$ damping; this is
the conventional weak-localizaton lineshape.
$B_1/B_{\phi} = 0$ corresponds to pure $\tau_3$ damping. The magnetic
field is in arbitrary units such that $B_{\phi} =1$.
We take $B_e = 200$. }

\newpage

\epsfxsize=3.0in \epsfbox{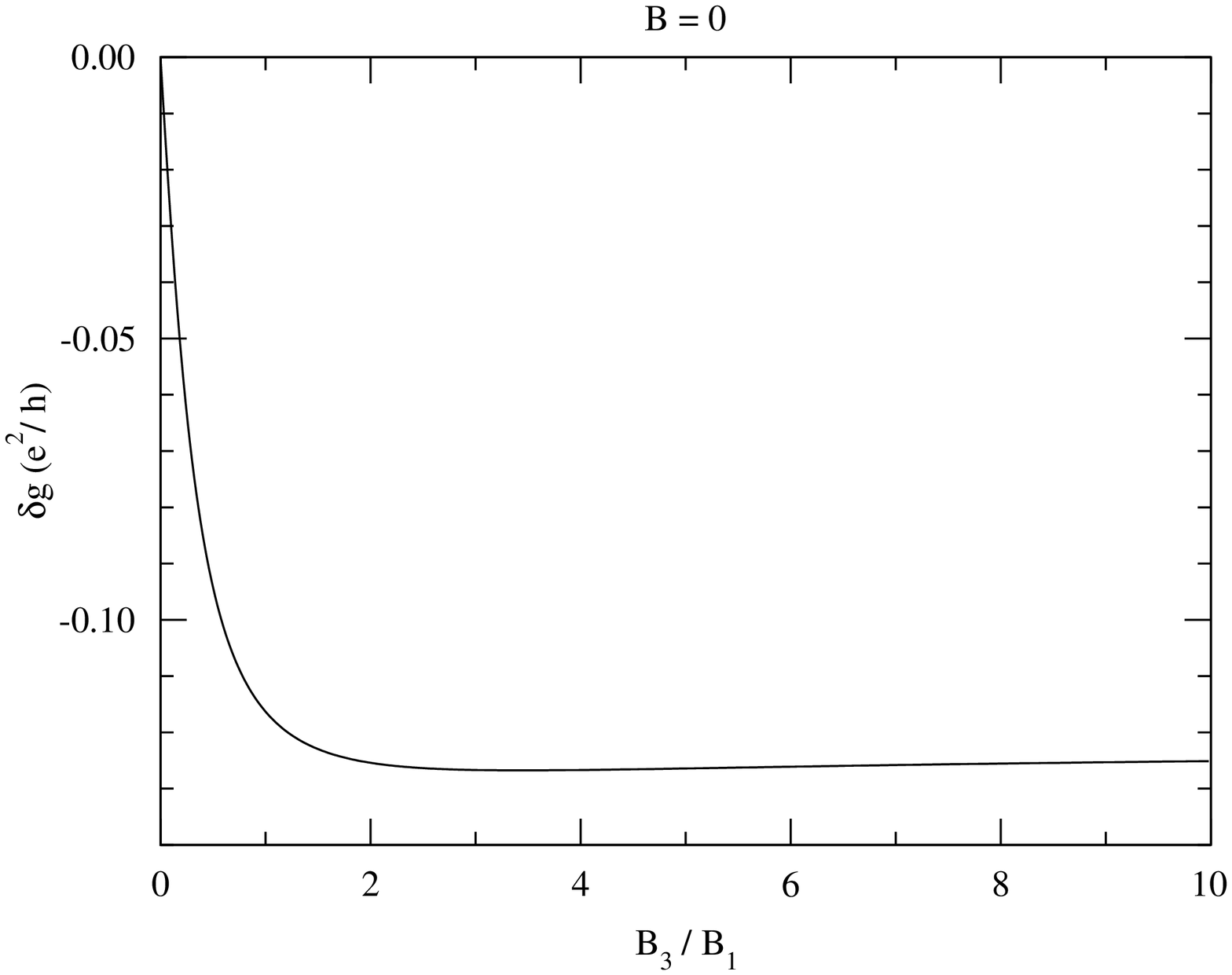}
\figure{ Figure 2. {\em Zero-field crossover:} The change in conductance
at zero field as the system varies from pure $\tau_1$ damping to
pure $\tau_3$ damping for a fixed total damping rate. The horizontal
axis is $B_3/B_1 = \tau_1/\tau_3$. $B_3/B_1 = 0$ corresponds to 
pure $\tau_1$ damping; $B_3/B_1 \rightarrow \infty$ corresponds
to pure $\tau_3$ damping. The vertical axis is the conductance
in units of $e^2/h$; the conductance at $B_3/B_1 = 0$ has been
subtracted.}


\end{document}